\documentclass[aps,pre,reprint,longbibliography,nofootinbib]{revtex4-1}

\usepackage{amssymb}
\usepackage{amsmath}
\usepackage{cases}
\usepackage{graphicx,color}
\usepackage{hyperref}
\definecolor{r}{rgb}{1,0,0}   
\definecolor{g}{rgb}{0,1,0}   
\definecolor{b}{rgb}{0,0,1}

\begin{document}


\title{Taut-Line Buzzers with Periodic Forcing}



\author{Jesse M. Hanlan$^1$, Douglas J. Durian$^{1,2}$}

\affiliation{$^1$Department of Physics and Astronomy, University of Pennsylvania, Philadelphia, PA 19104}

\affiliation{$^2$Department of Mechanical Engineering and Applied Mechanics, University of Pennsylvania, Philadelphia, PA 19104}


\date{\today}

\begin{abstract}
The time-dependent forcing and work per cycle required to drive sinusoidal spinning of a taut-line buzzer is analytically derived, both on and off resonance, from the nonlinear equation of motion. To test predictions, a model experimental system is constructed and characterized in terms of contraction versus twist angle and damped oscillations. The predicted force profile is then approximately implemented by hand. Nearly sinusoidal motion is observed, and the energy injection per cycle needed to maintain steady state oscillations is found to agree with theory. Additional force profiles are implemented, one in attempt to maximize non-sinusoidal response and one to minimize operator effort. With the latter, an Aluminum disk of radius 5~cm and height 0.95~cm was spun at a peak speed of over 11,000~RPM for fifteen minutes. The corresponding hand-powered centrifuge system would require $1.5\times$ more force, twice the power, and triple the time in order to run $90\times$ more samples than prior state-of-art.
\end{abstract}


\maketitle



\section{Introduction}
Sometimes deceptively simple machines can act as inspiration for technological advancement.
One such case is the ``buzzer" or ``whirligig" \cite{Schlichting2010}. This ancient toy is created by threading a string through a button or disk with two or more axisymmetric holes and leaving a loop of string on both sides. The disk is then rotated so the loops on each side become twisted; pulling on the loops causes them to rapidly untwist, imparting enough angular momentum to the disk that it can keep spinning and re-twists the loops in the opposite direction. Rhythmically pulling on the strings thus creates periodic twisting and untwisting, translating the linear pulling motion into very high RPM spinning motion of the disk. This conversion of low-frequency linear motion to high-frequency spinning motion is being investigated for hand-powered, low-cost centrifugation~\cite{Bhamla2017, Bhamla2019} as well as for portable energy generation and harvesting~\cite{Zhao2017, Tang2018, Chen2020}.

\begin{figure}[ht]
\includegraphics[width=3.2in]{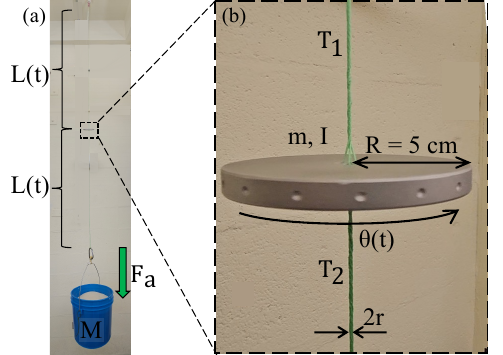}
\caption{Photos of a taut-line buzzer consisting of a spinning Aluminum disk of mass $m$ suspended between twisted Dyneema strings held under tension by a hanging mass $M$ to which a driving force $F_a$ is applied (positive if downward). The twisted strings, with tension $T_{1} = (M+m)g+F_{a}$ above and $T_{2} = Mg+F_{a}$ below, each exert a torque on the disk equal to their tension times the derivative $dL/d\theta$ of their length versus twist angle \cite{TautLineDamped}.}
\label{fig_setup}
\end{figure}

While the physics underlying the rich dynamics of hand-pulled buzzers is important to these efforts, it defies simple understanding and has previously required numerical analysis~\cite{Schlichting2010, Bhamla2017}. Furthermore, for the case of centrifugation, all rotational kinetic energy is lost when the disk stops spinning and must be totally re-injected by periodic pulling. Therefore, we are motivated to focus instead on an alternate ``taut-line buzzer" geometry introduced in Ref.~\cite{TautLineDamped}: a vertical implementation where the end of the top loop is fixed and the end of the bottom loop is attached to a hanging mass (Fig.~\ref{fig_setup}). We expect this geometry to enable easier and perhaps higher-speed operation of the buzzer by storing energy in the gravitational potential of the hanging mass, so that less energy injection is needed each cycle. An additional benefit is that an exact solution for the taut-line buzzer subject to periodic driving may be found, as demonstrated below. This helps inform physical understanding of the buzzer's dynamics and should aid in optimizing system parameters for applications.

This paper is organized as follows. In the next section, on theory, we review the equation of motion and its solution for damped oscillation, and then we develop exact solutions for the force needed to drive periodic sinusoidal motion both on and off resonance.  In the following section, on experiment, we present a large taut-line buzzer and characterize its natural oscillation frequency and $Q$ factor, and then we measure different force profiles and responses for periodic motion in order to compare with predictions.  Finally, we compare salient features of our driven buzzer with hand-pulled centrifuges and we discuss considerations for improving such applications.

\section{Theory of Taut-Line Buzzers}

\subsection{Equation of Motion}

In Ref.~\cite{TautLineDamped} we measured and modeled damped rotational oscillations for a series of taut-line buzzers such as depicted in Fig.~\ref{fig_setup}. There, we introduced these devices as a spinning mass $m$ with moment of inertia $I$ suspended between two inextensible but perfectly-flexible strings of effective radii $r$ and untwisted lengths $L_o$, kept under tension by a hanging mass $M$ to which an additional force $F_a$ may be applied.  When the disk is rotated by angle $\theta$, the strings above and below the disk twist in opposite directions and their resulting contractions cause the hanging mass to rise by distance
\begin{equation}
	y = 2(L_o-L) \approx (r\theta)^2/L_o,
\label{massheight}
\end{equation}
to leading order of contraction versus twist angle~\cite{TautLineDamped, Hanlan2023}. For this system, we established that the equation of rotational motion of the disk can be written as the sum of a spring-like linear restoring torque due to string tension and contraction versus twist angle plus a rate-dependent drag term due to viscous dissipation within the strings:
\begin{equation}
    I \ddot\theta = -2(T_o+F_a)\frac{r^2}{L_o} \theta - \frac{I\Omega_o}{Q}\dot\theta.
\label{EOM}
\end{equation}
Here, the average tension in the two strings for $F_a=0$, the natural oscillation frequency, and the $Q$ factor are respectively
\begin{eqnarray}
    T_o &=& (M+m/2)g \label{Tavg} \\
    \Omega_o &=& \sqrt{\frac{2T_o r^2}{IL_o}} \label{Omega} \\
    Q &\approx& \frac{\sqrt{2T_o I L_o}}{\eta_s r^3} \label{Qfactor} 
\end{eqnarray}
with $\eta_s$ being the string viscosity up to an unknown numerical prefactor \cite{TautLineDamped}. By convention the sign of $F_a$ is positive if the applied force is downward. In the equation of motion, $2(T_o+F_a)$ is the sum of tensions in the two strings, each of which must be positive. This limits the size of the upward force, $F_a<0$, that may be applied.

If, as can happen in experiment, the untwisted lengths $L_1$ above and $L_2$ below the disk are different, then the Eq.~(\ref{EOM}) of motion is generalized by replacing $2(T_o+F_a)/L_o$ with $(T_o+mg/2+F_a)/L_1 + (T_o-mg/2+F_a)/L_2 = [(M+m)g+F_a)]/L_1 + (Mg+F_a)/L_2$ \cite{TautLineDamped}. Often $m\ll M$ also holds, in which case the generalization is more simply $2/L_o \rightarrow 1/L_1+1/L_2$.

\subsection{Damped Oscillations}

In the absence of forcing, ordinary damped harmonic motion occurs where mechanical energy is traded back and forth between the disk (kinetic energy) and the hanging mass (gravitational potential energy), and is gradually dissipated by viscosity in the twisting strings \cite{TautLineDamped}. In particular, the equation of motion with $F_a=0$ gives the angular position of the disk and height of the hanging mass as
\begin{eqnarray}
    \theta(t) &=& \theta_o \cos(\Omega t+\phi)e^{-\Omega_o t/(2Q)} \\
    y(t) &=& y_o \cos^2(\Omega t+\phi)e^{-\Omega_ot/Q}
\end{eqnarray}
where $\Omega=\Omega_o\sqrt{1-1/(4Q^2)}$, $y_o=(r\theta_o)^2/L_o$, and the phase $\phi$ depends on the initial conditions. In the experiments below, the $Q$ values are large enough that $\Omega\approx\Omega_o$ holds to within experimental error, so the distinction is disregarded. Note that the potential energy is $(M+m/2)gy$ and the fractional energy loss per cycle is
\begin{equation}
    \Delta E/E = 2\pi/Q.
\label{losspercycle}
\end{equation}
In order to maintain steady oscillations, this loss must be replaced by work $W=\Delta E$ done by application of a driving force -- to be discussed next.

\subsection{Driving at Resonance}

One might despair of finding analytic solution for periodic forcing due to the nonlinearity arising from the product of $F_a$ and $\theta$ in the equation of motion. Indeed, we cannot solve for the motion that results from application of sinusoidal driving at the natural oscillation frequency $\Omega_o$. However, we may instead ask what applied force would be required to maintain periodic sinusoidal motion 
\begin{eqnarray}
    \theta(t) &=& \theta_o \cos \Omega_o t \label{qharmonic} \\
    y(t) &=& \frac{(r\theta_o)^2}{L_o}\cos^2\Omega_o t \label{yharmonic}
\end{eqnarray}
of the disk and hanging mass, respectively.  By direct substitution of this $\theta(t)$ into Eq.~(\ref{EOM}) it is possible to solve exactly for a required applied force of
\begin{equation}
    F_a(t) = \frac{T_o}{Q} \tan \Omega_o t
\label{Fa}
\end{equation}
This surprisingly simple form scales with $T_o$ as the characteristic force scale, and gets smaller at higher $Q$ as would be expected.  It is also distinctly not sinusoidal. In particular the required applied force diverges and changes sign at $\theta=0$ when the string is untwisted, at $y=0$ and maximum $\dot\theta$. In other words, the buzzer can be driven nearly sinusoidally by applying a down-then-up impulse to the hanging mass as it bottoms out and reached highest speed.

This prediction jibes with common experience operating a conventional hand-pulled buzzer, where the trick is to pull ever-harder as the disk accelerates while the string untwists, and then to suddenly stop pulling as soon as it's untwisted and begins to twist up in the opposite direction (for example see the force and speed versus time plots in Fig.2c of Ref.~\cite{Bhamla2017}). It's fun -and good exercise- to excite nearly-sinusoidal motion with a hand-pulled buzzer by maintaining an average pulling force $T_o$, on top of which a pull-relax impulse is applied at maximum extension. We discovered this empirically prior to the above analysis; however, to our knowledge, this way of driving a hand-pulled buzzer has not been mentioned in prior literature.


A curious feature of the above solution is that the amplitude $\theta_o$ of the driven rotations of the disk is undetermined.  Namely, multiple amplitudes of sinusoidal response can be achieved by the \textit{same} applied force, Eq.~(\ref{Fa}). As a check, we directly compute the injected work by integrating applied force versus hanging mass position, and the energy dissipated by integrating retarding torque versus disk angle, each over one cycle. Both are equal to
\begin{equation}
    W = 2\pi T_o y_o / Q
\label{work}
\end{equation}
where $y_o=(r\theta_o)^2/L_o$ is the maximum height reached by the hanging mass. Relatedly, the applied force can be expressed parametrically versus position as
\begin{equation}
    F_a(y) = \pm \frac{T_o}{Q}\sqrt{ \frac{y_o}{y}-1 }
\label{Fay}
\end{equation}
where $+$ is for pulling downwards as the hanging mass falls and $-$ is for pulling upwards as it rises. From integrating Eq.~(\ref{Fay}) versus $y$ one can see that most of the energy injection is for small $y$; \textit{e.g.} 50\% of energy injection is for $y<0.16y_o$ and 82\% is for $y<0.5y_o$.

\subsection{Driving off resonance}

A similar approach can be taken to finding the applied force required to drive the system sinusoidally at a frequency $\Omega \neq \Omega_o$ different from resonance,
\begin{eqnarray}
    \theta(t) &=& \theta_o \cos \Omega t \label{qoff} \\
    y(t) &=& \frac{(r\theta_o)^2}{L_o}\cos^2\Omega t \label{yoff}
\end{eqnarray}
Substituting $\theta(t)$ into the equation of motion, we exactly solve for a required force of
\begin{eqnarray}
    F_a(t) &=& F_o\frac{\sin(\Omega t + \delta)}{\cos\Omega t} \label{Faoff} \\
    F_o &=&  \frac{T_o}{Q}\frac{\Omega}{\Omega_o}\sqrt{1+\tan^2\delta} \label{Fooff} \\
    \delta &=& \tan^{-1}\left[ Q\left( \frac{\Omega}{\Omega_o}-\frac{\Omega_o}{\Omega} \right)\right] \label{doff} \\
\end{eqnarray}
Note that this correctly reduces to the on-resonance case for $\Omega \rightarrow \Omega_o$.  We will call $F_o$ and $\delta$ the ``amplitude" and ``phase", respectively, though these are misnomers compared to usual driven simple harmonic motion.


\begin{figure}
\includegraphics[width=3.2in]{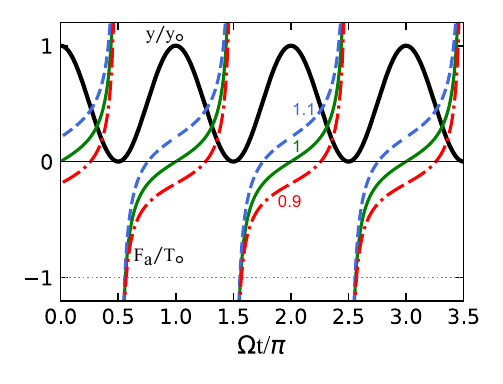}
\caption{Height $y(t)$ of hanging mass scaled by its maximum $y_o$ (thick black sinusoidal curve) and the Eq.~(\ref{Faoff}-\ref{doff}) prediction for the required downward applied force $F_a(t)$ scaled by average string tension $T_o=(M+m/2)g$ for different driving frequencies $\Omega/\Omega_o$ (as labelled: solid green at resonance; dashed blue above resonance; dot-dashed red below resonance) versus scaled time. These are for $Q=5$. Physically, the upward applied force cannot be less than the string tension ($F_a/T_o<-1$ is not allowed).}
\label{fig_VersusTime}
\end{figure}

To build intuition we now make a few instructive plots of the Eq.~(\ref{Faoff}-\ref{doff}) predictions. First, Fig.~\ref{fig_VersusTime} shows the applied force versus time in comparison with the height of the hanging mass, at resonance and at $\pm 10\%$ off, for $Q=5$. In all cases, the required force features a down-up impulse as the hanging mass bottoms out. The off-resonance cases appear qualitatively to be shifted up or down for respectively higher or lower frequencies. This jibes with the remarks above about exciting near-sinusoidal response in a hand-pulled buzzer, where the frequency is set by the average tension $T_o$ maintained in the strings.  Note that neither $F_a<-T_o$ nor the divergence of $F_a$ at $y=0$ are physically allowed. Therefore, in principle, perfectly sinusoidal motion cannot be excited in the laboratory. Nevertheless, nearly sinusoidal motion may be readily achieved -- as demonstrated in the experimental section below.

\begin{figure}
\includegraphics[width=3.2in]{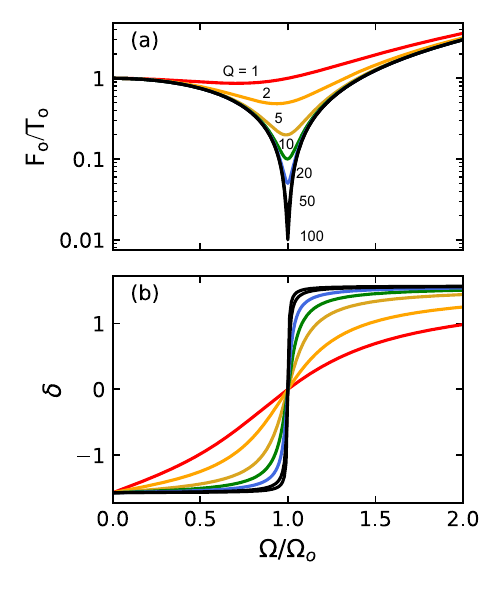}
\caption{Prediction of Eqs.~(\ref{Fooff}-\ref{doff}) for (a) the amplitude and (b) phase of the applied force versus frequency, for several $Q$ values. 
}
\label{fig_FaQvsW}
\end{figure}

Next, Fig.~\ref{fig_FaQvsW} shows the predicted amplitude and phase of the force needed for sinusoidal motion versus application frequency, for many $Q$ values. These are reminiscent of the familiar results for ordinary driven simple harmonic motion \cite{Marion}.  Namely, closer to resonance a smaller force is needed, which decreases with increasing $Q$. Versus frequency, the phase changes sign at resonance; this crossover, and the width of the resonance peak, both get sharper with increasing $Q$. Thus, the sinusoidal driving of taut-line buzzers conforms to much of our usual intuition for an oscillatory system even though it is intrinsically nonlinear and the applied force is highly non-sinusoidal.

\section{Experimental Demonstration}

We now test our theoretical predictions as follows.  First we describe the particular materials used in our experiment as well as the means of measuring both position and force. Next we characterize the effective twist radius of the string, and the unforced damping of the buzzer to find its natural frequency and $Q$ factor. We then demonstrate nearly-sinusoidal motion using the hand-powered approximation of the predicted force profile, as well as for two alternative force profiles.

\subsection{Apparatus}
The experimental system shown in Fig.~\ref{fig_setup} feature an Aluminum disk of radius $R=5.08$~cm, height $H=9.525$~mm, mass $m=215.8$~g, and moment of inertia $I=2780~\textrm{g}\cdot\textrm{cm}^2$ (disk ``G" from \cite{TautLineDamped}). This is suspended using a 12-strand ultra-high-molecular-weight polyethylene (UHMWPE) Dyneema line (Emma Kites) of nominal radius $r=0.4$~mm and quoted maximum breaking strength of 220~pounds (980~N). A 12~m length of this string is looped through four holes drilled in a square pattern symmetrically around the buzzer center and knotted together at the ends, forming two loops on each side that are then loosely hand-braided to prevent the buzzer from sliding while hanging~\cite{twoloopvideo}. This type of line is the same as used by Bhamla \textit{et al.}~\cite{Bhamla2017}, and is stronger than the Nylon line used in our damping experiments~\cite{TautLineDamped} so that here we can minimize the string diameter. Whereas Refs.~\cite{Bhamla2017, TautLineDamped} respectively use a two-strand loop and single strand, the use here of loosely-braided 4-string bundle for the buzzer has several benefits: The rotation is more consistently on axis since the setup is axisymmetric (no knots near the disk), the working load limit is increased by distributing the hanging mass over four strands, there is no energy loss from the strands flying apart when untwisted at high speed, and the redundancy in strings reduces the risk of injury should one strand break during operation. Indeed, in early trials the failure mode was always for one strand to snap but the other three prevented the hanging mass and buzzer from crashing to the floor.  After braiding, the top string is attached to the ceiling by a carabiner and an approximately 20~kg bucket of sand is hung from the bottom string. For the data shown below, the string on either side of the buzzer has an uncontracted length of $L_o=1.5$~m. The position of the hanging mass is measured using a range finder (Vernier, Motion Detector 2 and Logger Pro) from the floor to the bottom of the hanging mass.

\begin{figure}
\includegraphics[width=2.5in]{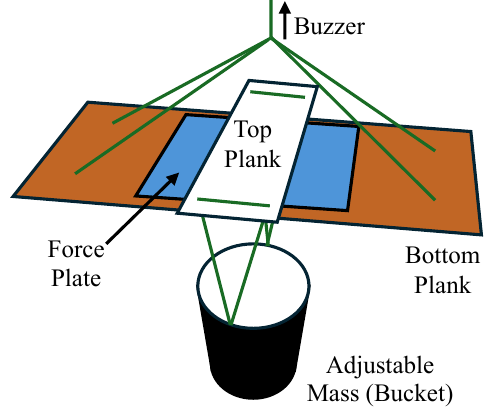}
\caption{Force plate sensor, wood planks, and adjustable hanging mass used for measuring force data during buzzer driving. The force plate (black) is under compression between the bottom wood plank (brown) attached to the buzzer and the top wood plank (white) from which the adjustable mass bucket is hanging. During driving, force is applied to the top plank with a maximum upward applied force equal to the mass of the top plank and bucket.
}
\label{fig_ForcePlate}
\end{figure}

In order to measure the applied force, a force sensor (Vernier, Force Plate and Logger pro) is placed in-line with the hanging mass as shown in Figure~\ref{fig_ForcePlate}. The sensor measures compressive force, requiring a two stage apparatus. The force plate rests on a wood plank hung from a carabiner attached to the bottom string of the buzzer. A second wood plank is placed on top of the force sensor such that it overhangs on the sides; a bucket of sand is hung from the top wood plank and hangs freely below the bottom wood plank. Thus the force sensor begins in a state of compression, supporting the weight of the top plank and bucket and resting on the bottom plank hanging from the buzzer bottom string. Upward force applied to the top plank can be measured as this compression is mitigated, as long as the top plank remains in contact with the force plate. Our maximum applied force is limited by the force compressing the force plate, $(M_{\text{plank}} + M_{\text{bucket}})g$ or approximately $200$~N. This differs from the effective mass hanging from the bottom string; the total hanging mass includes both planks of wood, force plate, and filled bucket and is tuned to equal $M=25$~kg. This corresponds to 55~lbs, safely below the maximum breaking strength of $4\times200$~lbs.

\subsection{Contraction and Damped Oscillation}
\begin{figure}[ht]
\includegraphics[width=3.2in]{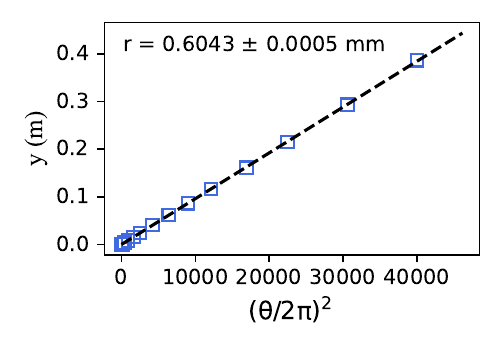}
\caption{Height of hanging mass above its low point versus square of twist angle for the taut-line buzzer set-up depicted in Fig.~\ref{fig_setup}. The line represents a fit of the data to Eq.~(\ref{massheight}), which gives the effective twist radius for the loosely braided loops as $r = 0.6043\pm 0.0005$~mm.}
\label{fig_twist}
\end{figure}

Because we do not measure the disk angle directly, we need to convert the hanging mass position into twist angle via testing and calibrating with Eq.~(\ref{massheight}). As was shown in Ref.~\citenum{Hanlan2023}, the nominal / geometric strand radius cannot be used directly, though the radius of gyration for a bundle provides an effective proxy~\cite{Hanlan2023}. For a four-strand bundle this defined by its moment of inertia as $I_\mathrm{bundle} =(4m)r_\mathrm{gyr}^2$, which gives $r_\mathrm{gyr}$ as $\sqrt{5/2}$ times the geometric radius of the strands whether they are arranged as a square or a diamond. For strands of radius $0.4$~mm, we therefore expect the effective twist radius to be $r = r_\mathrm{gyr} = 0.6325$~mm. To find the true value of the effective twist radius, we follow the procedures of Ref.~\cite{Hanlan2023} by measuring the height of the hanging mass above bottom and plotting it versus the squared twist angle in Figure~\ref{fig_twist}. We find the contraction collapses well to the linear expectation set by Eq.~(\ref{massheight}); fitting this data to a line gives an effective radius for the string as $r = 0.6043 \pm 0.0005$~mm, less than $5\%$ different from expectation.

\begin{figure}[ht]
\includegraphics[width=3.2in]{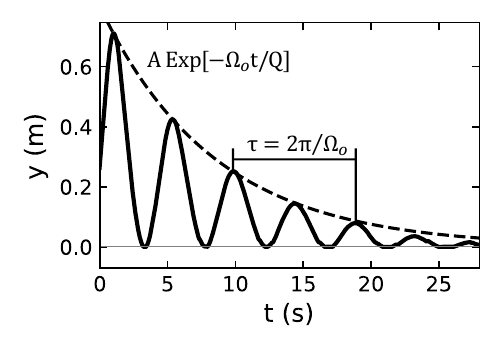}
\caption{Height vs time data for damped oscillations, without external forcing, measured for the set-up depicted in Fig.~\ref{fig_setup}. The time between every other peak gives a period of $\tau=8.861 \pm 0.0002$~s and natural frequency $\Omega_{o} = 0.709 \pm .006$~rad/s.  The exponential envelope gives $Q = 5.9 \pm 0.1$.}
\label{fig_damping}
\end{figure}

We now similarly follow Ref.~\cite{TautLineDamped} in order to characterize damped oscillations in the absence of forcing. In particular, the buzzer is set to a large twist angle and allowed to spin without forcing until the oscillations naturally cease. We record the vertical position of the hanging mass versus time and report it in Figure~\ref{fig_damping}. From this we directly measure the period from the time between alternate peaks, giving a natural frequency $\Omega_o = 0.71 \pm 0.01$~rad/s similar to the predicted value 0.65~rad/s from Eq.~(\ref{Omega}).

\begin{figure}
    \centering
    \includegraphics{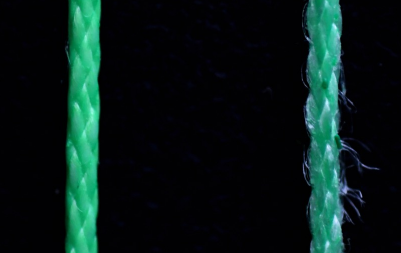}
    \caption{Dyneema string of nominal radius $r=0.4$~mm before (left) and after (right) several runs of the taut-line buzzer.}
    \label{fig:StringDamage}
\end{figure}

Next we fit the peaks of the damped oscillations to an exponential envelope of the form $y \propto \exp(-\Omega_o t/Q)$. This fit gives the quality factor as $Q=5.9\pm 0.1$. In turn, Eq.~(\ref{Qfactor}) then gives an effective string viscosity of $\eta_s = 3.47 \times 10^{9} ~\mathrm{g/(cm \cdot s)}$. This is $75\times$ greater than that of the single-stranded Nylon rattail cord used in Ref.~\cite{TautLineDamped}. Visual inspection reveals an accumulation of damage over the course of use, per photos in Fig.~\ref{fig:StringDamage}. The strings, initially green, lose much of their hue and green flecks can be found on top of the force plate. Additionally, a looser single-strand braid structure and small broken filaments extending radially outward from the surface can be seen in the image. It is unclear whether this damage is due to gradual plastic failure (strings can be broken by over-twisting) or to rubbing between fibers or between strands. Damage accumulation did not noticeably affect behavior, but can be expected to eventually lead to failure.  Since no such damage was ever noticed with the use of Nylon rattail strings in Ref.~\cite{TautLineDamped}, that type of line seems better suited for applications.

As a final note on characterizing the setup, we note that $4Mr^2=360$~g$\cdot$cm$^2$ is small compared to $I=2780$~g$\cdot$cm$^2$.  Therefore, according to the criteria developed in Ref.~\cite{TautLineDamped}, the translational kinetic energy of the hanging mass may be neglected in the equation of motion even at the highest twist angles and speeds.

\subsection{Periodic Forcing}

We now attempt to apply the time-dependent driving force predicted to give periodic sinusoidal angular motion. The required force profile shown parametrically in Eq.~(\ref{Fay}) presents a number of experimental challenges. As the hanging mass reaches its minimum height, the magnitude of the force diverges and the direction instantly changes, both of which are impossible to replicate manually. If the sign change of the applied force is a little late, then it removes rather than injects energy.  As noted above, the force plate only measures compressive force and thus we are limited to a maximum upward force of approximately 200~N. To minimize asymmetry in the motion, we impose the same limit on the downward force. The increase in force happens over a small distance, with 50\% of the work injected into the system in the bottom 16\% of height; in our apparatus this translates to $\approx 5$~cm. 

\begin{figure}[ht]
\includegraphics[width=3.2in]{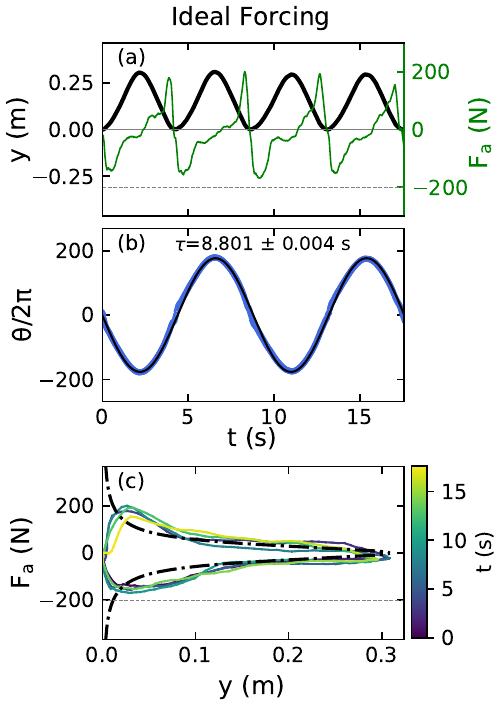}
\caption{Dynamics data for `ideal' forcing, as reasonably close as can be achieved by hand to the Eq.~(\ref{Fa}) prediction needed for perfectly sinusoidal motion: (a) Directly-measured height (left, thick black) and applied force (right, thin green) and (b) inferred twist angle vs time. In (a), the maximum measurable upward force is represented as a horizontal dashed gray line. In (b), the expected sinusoidal form for $\theta(t)$ is fit in black, giving the period as shown. (c) Force vs height plotted parametrically over the same time range as (a-b); the measured force profile is in green, and the predicted ideal force profile of Eq.~(\ref{Fay}) is in black.
}
\label{fig_ideal}
\end{figure}

Despite these challenges, with some practice we were able to achieve a reasonably close by-hand approximation of the forcing profile needed for ideally sinusoidal motion. Steady operation with nearly-sinusoidal response is demonstrated in a supplemental video~\cite{tautlinevideo}.  More quantitatively, in Fig.~\ref{fig_ideal} we display the directly-collected position (black) and force (green) data vs time for two representative periods. The maximum allowed upward force is shown as a dotted gray line ($F_a=-200$~N, recalling the convention that positive $F_a$ is for a downward applied force). The angular position, using the conversion from Fig.~\ref{fig_twist}, is shown in Fig.~\ref{fig_ideal}b with raw data in blue and the theoretical prediction in black. We see that the motion is indeed highly sinusoidal and closely matches the natural frequency; namely, the forcing frequency, $\Omega = 0.7149$~rad/s, is less than 1\% different from the value $\Omega_{o} = 0.7091$~rad/s measured from damped oscillations. In Fig.~\ref{fig_ideal}c, we see the applied parametric force profile fairly matches with theory, despite the challenges outlined above. By integrating Fig.~\ref{fig_ideal}c, we find the total work applied over one period to be 82~J per cycle. We can use this applied work and Eq.~(\ref{losspercycle}) in order to define the quality factor of this driving: $Q=2\pi E/W = 5.8 \pm 0.2$. This value closely matches the quality factor measured in the damped oscillator, $Q = 5.9 \pm 0.1$.

\begin{figure}[ht]
\includegraphics[width=3.2in]{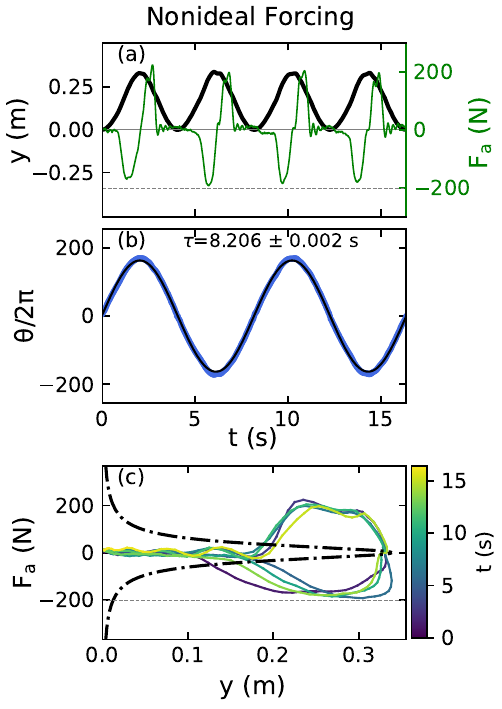}
\caption{Dynamics data for a force profile which is `nonideal' in being of opposite magnitude to the Eq.~(\ref{Fa}) prediction for perfectly sinusoidal motion: (a) Directly-measured height (left, thick black) and applied force (right, thin green) and (b) inferred twist angle vs time. In (a), the maximum measurable upward force is represented as a horizontal dashed gray line. In (b), the expected sinusoidal form for $\theta(t)$ is fit in black, giving the period as shown. (c) Force vs height plotted parametrically over the same time range as (a-b); the measured force profile is in green, and the predicted ideal force profile of Eq.~(\ref{Fay}) is in black.
}
\label{fig_InefficientForcing}
\end{figure}

For comparison, we now present a `nonideal' force profile, which is large when the `ideal' case is small and vice-versa. Namely, as shown in Figure~\ref{fig_InefficientForcing}, we apply the strongest impulse at peak contraction, rather than peak velocity, and observe the resulting behavior.  This should give a response that is highly non-sinusoidal. We find that the peak heights are similar, $y_o = 0.310~m$ (ideal) and $y_o = 0.338~m$ (nonideal) for similar maximum applied forces $F_o = 201~N$ (ideal) and $F_o = 223~N$ (nonideal). Looking at Fig.~\ref{fig_InefficientForcing}b, we surprisingly find that the resulting rotational motion is approximately as sinusoidal as the ideal case.  Thus, the taut-line buzzer geometry is remarkably robust in giving a nearly-sinusoidal response for a broad range of periodic force profiles. The primary difference we observe between the ideal and nonideal cases is their efficiency.  Namely, the quality factor for the `nonideal' case, $Q = 5.5 \pm 0.2$, is lower than that of the `ideal' case above. These and other metrics for all force profiles, including the unforced case, are presented in Table~\ref{DrivingMetrics}.

\begin{figure}[ht]
\includegraphics[width=3.2in]{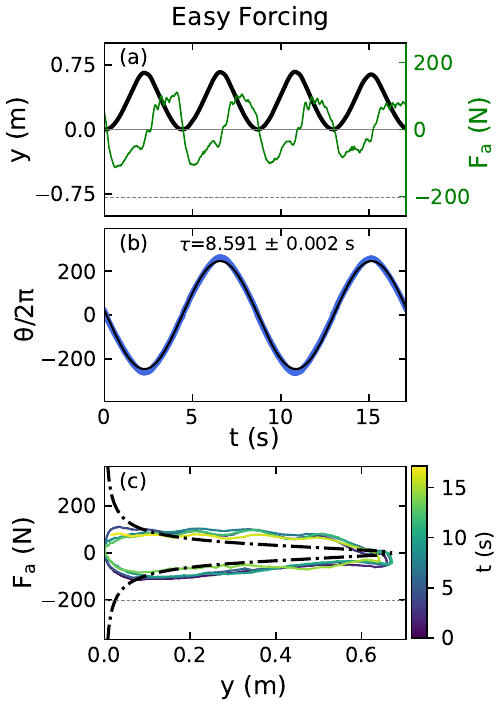}
\caption{Dynamics data for a force profile that we find natural and `easy' to apply by hand: (a) Directly-measured height (left, thick black) and applied force (right, thin green) and (b) inferred twist angle vs time. In (a), the maximum measurable upward force is represented as a horizontal dashed gray line. In (b), the expected sinusoidal form for $\theta(t)$ is fit in black, giving the period as shown. (c) Force vs height plotted parametrically over the same time range as (a-b); the measured force profile is in green, and the predicted ideal force profile of Eq.~(\ref{Fay}) is in black.
}
\label{fig_EasyForcing}
\end{figure}

By comparing the `ideal' and `nonideal' cases, we note that the two force profiles are differentiated primarily by the effort required to achieve steady-state motion. Therefore, we now investigate an alternative force profile that minimizes the difficulty required for steady operation. By feel and experience, we tend to naturally settle on an `easy' force profile as illustrated in Fig.~\ref{fig_EasyForcing}.  This resembles a square wave with smoothed transitions the height extrema -- which is much easier to achieve by hand than the sudden down/up impulse demanded for the `ideal' case.  As seen in Fig.~\ref{fig_EasyForcing}a, the resulting peak height of the hanging mass during `easy' driving is over twice that of the `ideal' case. This converts to a roughly $1.5\times$ increase in the peak twist angle, shown in Fig.~\ref{fig_EasyForcing}b, and corresponds to a maximum angular velocity of over $11,000$~RPM. The angular data appears sinusoidal, though we note systematic overestimation in the fit at medium twist angles and underestimation at maximum twist, which demonstrates the motion is only sinusoidal in approximation.

\begin{table*}
\caption{\label{DrivingMetrics}%
Taut-line buzzer metrics for damped motion and for the three force profiles. The period is $\tau$, the maximum rise height of the hanging mass is $y_o$, the maximum angular speed is $\omega_o$, the maximum radial acceleration at the edge of the disk is $a_o$, the injected work per cycle is $W$, the quality factor is $Q$, and the maximum applied force is $F_o$.
}
\begin{ruledtabular}
\begin{tabular}{lccccccc}
Experiment & $\tau$~(s) & $y_o$~(cm) & $\omega_o$~(RPM) & $a_o/g$ & $W$~(J) & $Q$ & $F_o$~(N) \\
\colrule
Damped & 8.861 $\pm$ 0.002 & 71.1 $\pm$ 0.5 & - & - & - & 5.9 $\pm$ 0.1 & - \\
Ideal & 8.801 $\pm$ 0.006 & 31.0 $\pm$ 0.5 & 7,540 $\pm$ 70 & 3180 $\pm$ 30 & 82 $\pm$ 4 & 5.8 $\pm$ 0.2 & 201 $\pm$ 3 \\
Nonideal & 8.206 $\pm$ 0.004 & 33.8 $\pm$ 0.5 & 7,510 $\pm$ 70 & 3155 $\pm$ 30 & 95 $\pm$ 4 & 5.5 $\pm$ 0.2 & 223 $\pm$ 3\\
Easy & 8.588 $\pm$ 0.003 & 67.3 $\pm$ 0.5 & 11,150 $\pm$ 70 & 6830 $\pm$ 40 & 181 $\pm$ 4 & 5.7 $\pm$ 0.2 & 113 $\pm$ 3 \\
\end{tabular}
\end{ruledtabular}
\end{table*}

In Fig.~\ref{fig_EasyForcing}c, we characterize the ease of this force profile in terms of the peak force and injected work work. Compared to the peak force in the `ideal' case, $F_o = 201$~N, the `easy' driving peak force is nearly halved, $F_o=113$~N. Despite the substantially lower peak force, the `easy' driving injected over twice the work per cycle, $W=181~\mathrm{J}$ (easy) and $W=82$~J (ideal). The quality factor for this profile, $Q = 5.7 \pm 0.2$, shows that it is marginally less efficient than the `ideal' case; however, given the substantial increase in peak velocity for smaller peak force we consider this to be an negligible cost.

Since the ideal/non-ideal/easy force profiles all lead to periodic motion that appears highly sinusoidal, we explore a little further the robustness of such behavior in the Appendix.

\section{Hand-powered Centrifugation}

In this final section we consider the possible use of taut-line buzzers for hand-powered centrifugation. We begin by comparing the above taut-line buzzer system with prior hand-pulled systems.  Then we consider how the parameters of our set-up might be varied in order to improve performance.

\subsection{Taut-line versus Hand-pulled}

Since our Aluminum disk has the same radius as the hand-pulled ``Paperfuge" disk of Ref.~\cite{Bhamla2017}, we can make some direct comparisons as though our disk were replaced by a mass-equivalent number of stacked Paperfuge samples. Based on the masses and maximum speeds listed in Table~\ref{CentMetrics}, our Fig.~\ref{fig_setup} taut-line buzzer with `easy' driving corresponds to 90 Paperfuge samples driven at about half the speed. The average radial acceleration at the disk edge is a convenient metric for rate of centrifugal sedimentation \cite{Bhamla2017}.  This quantity is $\langle a\rangle = \omega_o^2R/2$ in our case, but is reduced by more than $1/2$ from $\omega_o^2R$ for the Paperfuge because its motion is nonsinusoidal. As listed in Table~\ref{CentMetrics}, we achieve $\langle a\rangle/g = 3480$ while the Paperfuge achieves $\langle a\rangle/g = 9499$.  Thus, our present setup would require nearly $3\times$ longer to run $90\times$ more samples. This is certainly feasible: As a useful as well as healthful exercise, we ran our taut-line buzzer for 15~minutes without undue effort.

\begin{table*}
    \caption{\label{tab:PaperfugeComparison} Centrifugation metrics for the Aluminum disk taut-line buzzer with `easy' driving (this work), the Paperfuge~\cite{Bhamla2017}, and the 3D printed centrifuge (3D-Fuge)~\cite{Bhamla2019}. }
\begin{ruledtabular}
\begin{tabular}{cccccccccc}
         Buzzer & $m$~(g) & $R$~(cm) & $I$~(g$\cdot$cm$^2$) & $\omega_o$~(RPM) &  $\mathrm{\langle a \rangle}$/g & $K_o$~(J) & $\tau$ (s) & $F_o$ (N) & $P$ (W)\\
 \hline
 Taut-Line & 215.8 & 5 & 2780 & 11,150 & 3480 & 190 & 8.6 & 113 & 21 \\
 Paperfuge & 2.4 & 5 & 30 & 22,065 & 9499 & 8.0 & 2 & 75 & 8 \\
 3D-Fuge \#2 & 13.5 & 4.5 & 137 & 6,000 &  1000 & 2.7 & 1.5 & unknown & 3.6 \\
\end{tabular}
\end{ruledtabular}
\label{CentMetrics}
\end{table*}

In addition to run times, we should also consider ease of use.  One metric for this is the maximum applied force, which scales as $F_o=T_o/Q$.  This is $F_o=113$~N in our case and 75~N for the Paperfuge case. Another metric is the average rate of energy injection. In our case this is $P=W/\tau=21$~W, the total injected energy per period.  For the Paperfuge, this may be estimated as $P=2K_o/\tau=8$~W because the maximum kinetic energy is lost and re-injected twice per period. However, this is an underestimate since it ignores dissipation during the driving process. Note that the ratio of powers is greater than the ratio of maximum forces because of the differing force profiles and ranges of motion as set by untwisted string lengths ($L_o$=1.5~m for the taut-line buzzer, $L_o$=0.33~m for the Paperfuge). Altogether, a centrifugation system equivalent to our taut-line buzzer would thus require roughly $1.5\times$ more force, twice the power, and triple the time in order to run $90\times$ more samples than the Paperfuge. 

A similar comparison can also be made to the ``3D-Fuge" design \#2 of Ref.~\cite{Bhamla2019}. For that setup, the force profile and moment of inertia are unknown; for the latter, we approximate the spinning mass as a solid disk. Based on the numbers given in Table~\ref{CentMetrics}, a centrifugation system equivalent to our taut-line buzzer would thus require roughly $5\times$ more power and one third the time in order to run $16\times$ more samples than the 3D-Fuge.

There are several additional regards in which the taut-line buzzer approach may be helpful for centrifugation applications.  First, its vertical geometry ensures that all translational motion of the hanging mass and the disk is co-linear with the rotation axis. By contrast, the hand-pulled Paperfuge and 3D-Fuge geometries are horizontal and the disks necessarily bob up and down, perpendicular to the rotation axis, when pulled.  This is a waste of kinetic energy, and also tends to excite transverse waves in the string that can further enhance dissipation.  Second, the taut-line buzzer has a natural frequency and therefore the user has no choice but to drive the system at this required periodicity.  By contrast, the hand-pulled buzzer is brought to a full stop each half period and the user must take initiative to pull as soon as possible after the spinning stops. Users getting bored or tired may tend to pause to varying degrees before the next pull. Thus, the taut-line buzzer is more amenable to minimizing the operating time for a given set-up and to maximizing its reproducibility. Third, the hand-pulled buzzer requires the user to exert a force by pulling their hands apart.  This is an unnatural motion for exerting large forces, unlike rowing a boat or peddling a bike. By contrast, for the taut-line buzzer, the user is free to employ a more ergonomic lifting motion or, for example, to rig up a hand- or foot-powered lever to force the hanging mass. Lastly, since the maximum untwisted string length is not limited by the human wingspan, it may be significantly increased in order to optimize operation -- as considered next.

\subsection{Optimizing Experimental Parameters}

For speed and ease of operation, the goal is to maximize both the average radial acceleration at the edge of the disk and the quality factor for the oscillations, and relatedly to minimize the scale $T_o/Q$ of the applied force.  The former is $\langle a\rangle = \langle \omega^2 R\rangle = \omega_o^2R/2$ per above, which may be estimated more generally as follows. First note that the maximum angular speed of the spinning disk is set by Eq.~(\ref{qharmonic}) as $\omega_o=\theta_o\Omega_o$, where $\theta_o$ is the maximum twist angle and $\Omega_o$ is the driving frequency. If the system is driven at maximum amplitude with no supercoiling, then $\theta_o=c(L_o/r)$ holds where $c$ is a materials-dependent numerical factor typically in the range 0.8--0.9~\cite{Hanlan2023}. Using Eq.~(\ref{Omega}) and assuming the spinning mass is a disk with $I=(1/2)mR^2$ then gives
\begin{equation}
    \langle a\rangle < \frac{1}{2}\left[c(L_o/r)\Omega_o\right]^2R = \frac{2c^2 T_oL_o}{mR},
\label{aavg}
\end{equation}
which naively suggests maximizing $\{T_o,L_o\}$ and minimizing $\{m,R\}$ separately. However, there are several practical considerations that are not obvious how to accommodate.  First, the quantities in Eq.~(\ref{aavg}) affect $\Omega_o$ and it is not humanly reasonable to drive the system with a period shorter than about $\tau=1$~s, \textit{i.e.}~with a frequency greater than about $\Omega_o=2\pi$~rad/s. Second, $R$ should be at least twice as large as the size of the specimen being spun down. Third, $T_o$ plus driving force should be kept below a safety factor $s$ times the maximum breaking strength of the string, $\sigma_y \pi r^2$ where $\sigma_y$ is the tensile yield stress.  For the taut-line buzzer system above, the safety factor is $s=0.06$. The value of $\sigma_y$ depends on material and braid structure; for Dyneema it is about $\sigma_y=2000$~MPa = 450~lb/mm$^2$, for Nylon rattail it is about ten times smaller.  Using $T_o<s\sigma_y \pi r^2$ and $m=\rho \pi R^2H$ gives the maximum possible average radial acceleration and quality factor as
\begin{eqnarray}
    \langle a\rangle &<& \frac{2c^2s\sigma_y L_o r^2}{\rho H R^3} \\
    Q &<& \frac{\sqrt{\pi s\sigma_y\rho H L_o} R^2}{\eta_s r^2}
\end{eqnarray}
For quicker and easier operation, one should clearly choose strings that are longer, stronger, and less dissipative. Since the other parameters have opposite effects on speed and ease of operation, one must compromise and decide to what extent $\langle a\rangle$ or $Q$ is more important to optimize.  However, the height $H$ is actually an exception -- the acceleration goes down but the number of samples goes up in same proportion and hence the net effort per sample is unchanged. Since larger $H$ gives larger $Q$, it is advantageous as long as $\Omega$ does not become so low that the system is difficult to drive by hand.  

\begin{figure}[ht]
\includegraphics[width=3.2in]{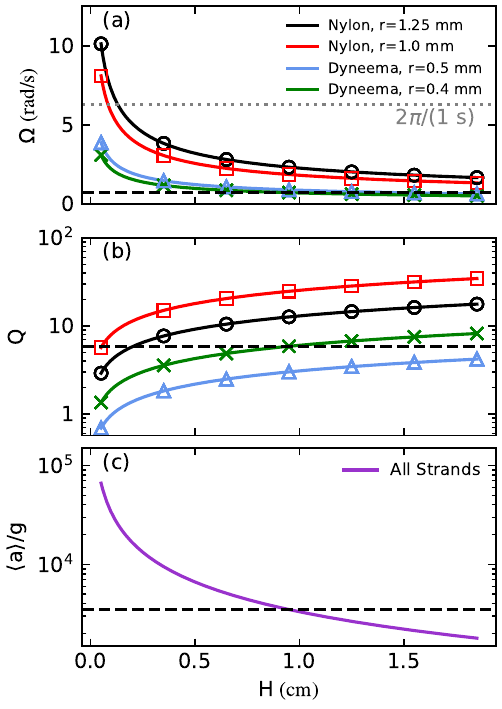}
\caption{Predictions for the (a) natural frequency, (b) $Q$ factor, and (c) average acceleration based on scaling the measured values (dashed lines) according to Eqs.~(\ref{Omega},\ref{Qfactor},\ref{aavg}) if different strings and Aluminum disks of different heights $H$ are swapped into the existing setup ($r=0.4$~mm Dyneema with $H=0.9524$~cm). The curves are cut off for $H<0.05$~cm where either the natural frequency is too great to drive by hand or $Q$ is too small. }
\label{fig_Optimization}
\end{figure}

To illustrate this trade off, we now predict possible improvements upon our experimental taut-line buzzer setup. Since the values of $L_o$ and $T_o$ are already as large as can be readily accommodated, we keep these fixed. We also keep fixed the material and radius of the disk. Starting from measured values and scaling by Eqs.~(\ref{Omega}, \ref{Qfactor}, \ref{aavg}), the predicted natural frequency, quality factor, and average acceleration are plotted in Fig.~\ref{fig_Optimization} versus disk height $H$ for four possible string materials and radii. The 1.25~mm radius Nylon rattail (black) has the same $s=0.06$ safety factor as the currently used 0.4~mm radius Dyneema, but has a viscosity that is $\times75$ smaller. The smaller radius Nylon (red) has marginally worse safety factor ($s=0.098$) whereas the larger radius Dyneema (blue) is safer ($s=0.04$).

We first note from Fig.~\ref{fig_Optimization}a that all but the thinnest disks using Nylon string have a natural frequency less than $2\pi$ (gray dotted line), meaning virtually the whole given range can be driven by hand. We see the immediate improvement from switching to Nylon in Fig.~\ref{fig_Optimization}b where the quality factor of the current apparatus is represented as the black dashed line. Nearly the entire range of Nylon string data, regardless of radius, is above the current quality factor, indicating the Nylon as a strict improvement over the Dyneema for energy conservation between cycles. The 1.25~mm radius Nylon, having the same safety factor as the Dyneema above, represents a safe improvement while the 1.0~mm radius Nylon could offer still greater benefits with some safety testing. Finally we see the average acceleration, Fig.~\ref{fig_Optimization}c, is identical regardless of string radius or material, and thus represents a design choice. 

Lastly, as one possible figure of merit, we consider maximizing the effective amount $\langle a\rangle H$ of centrifugation per scale $T_o/Q$ of applied forcing. The ratio of these quantities scales as $\sqrt{T_o}$, suggesting it's good to maximize the hanging mass to its safe limit $T_o=s\sigma_y\pi r^2$. Using this with Eqs.~(\ref{Omega},\ref{Qfactor},\ref{aavg}) gives the figure of merit as
\begin{equation}
    \mathcal{C/F} = \frac{\langle a \rangle H}{T_o/Q} = 
    \frac{2c^2\sqrt{s\sigma_y}L^{3/2}}{\eta_s r^2}~\frac{H^{1/2}}{\sqrt{\rho}R}
\label{figureofmerit}
\end{equation}
where the first term pertains to the string and the second to the disk.  For a given string material, longer and thinner is better.  Compared to Dyneema, the Nylon rattail is better by a factor of $\sqrt{1/10}/(1/75)\approx 25$ because it is $\times 10$ less strong but $\times 75$ less dissipative.  As for the disk, it's good to minimize its density and take the radius as not much larger than the sample length. Running multiple samples (large $H$) is also good for increasing $\mathcal{C/F}$.  All these desirables, except for minimizing $\rho$ and $R$, cause a decrease in $\Omega$.

\section{Conclusion}
We characterized the response of the taut-line buzzer introduced in Ref.~\cite{TautLineDamped} to external forcing. We showed that, while the governing equation~\ref{EOM} is nonlinear, one can nevertheless identify the required force profile for sinusoidal disk motion. We performed such forcing by hand, and observed the resulting  motion. When compared to alternate forcing profiles, we found the motion still appeared highly sinusoidal, however the work required to drive the system increased relative to its motion. We also characterized the force profile that was easiest to operate and elicited the greatest driving response. The rotation of the disk under this `easy' force profile was compared to the paper centrifuge (Paperfuge) described in Ref.~\cite{Bhamla2017}. We found the taut-line aluminum buzzer was the equivalent of spinning a stack of 90 Paperfuges without significantly greater effort. In particular, we demonstrated that such a large platform can be driven by hand for extended periods by a single operator, presenting a possibility to centrifuge many milliliters of sample at a time in areas without access to electricity or expensive equipment. Finally we identified the physical constraints on some system variables such as forcing period and string length. We predicted how changing string radii, material, and disk height would impact the key metrics for centrifugation and discussed the trade-offs one might consider in optimizing their own taut-line buzzer depending on centrifugation protocol.


\begin{acknowledgments}
We thank Peter Harnish and Mary Marcopul for instrumentation support. This work was partially supported by NSF grants MRSEC/DMR-1720530 and MRSEC/DMR-2309043. DJD thanks CCB at the Flatiron Institute, a division of the Simons Foundation, as well as the Isaac Newton Institute for Mathematical Sciences under the program ``New Statistical Physics in Living Matter" (EPSRC grant EP/R014601/1), for support and hospitality while a portion of this research was carried out.
\end{acknowledgments}


\appendix
\section{Non-Sinusoidal Motion}

\begin{figure}[ht]
\includegraphics[width=3.2in]{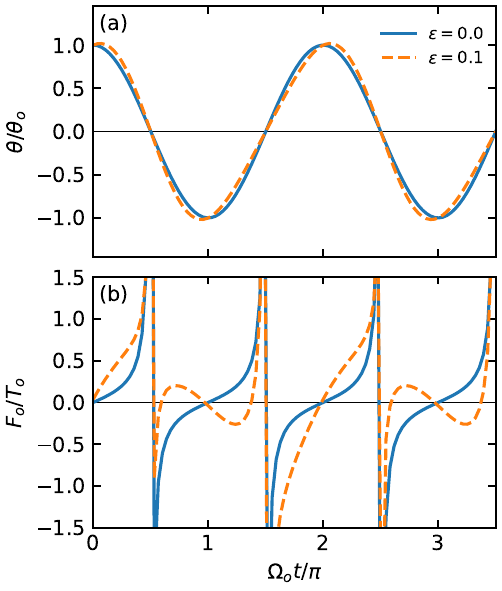}
\caption{Predictions of Eqs.~(\ref{eq_anharmonicmotion},\ref{eq_anharmonicforce}) for (a) the periodic motion resulting from (b) two different time-dependent applied force profiles.  In both cases the quality factor is $Q=5$.
}
\label{fig_appendix}
\end{figure}

The experimental observation that three driving force profiles differing variously from Eq.~(\ref{Fa}) all lead to responses that are nearly sinusoidal raises the question of how robust is such behavior. Perhaps this could be analyzed in general or by numerical solution for the shape of the response to various applied forcing profiles. For such work, it may be useful to note that the equation of motion~(\ref{EOM}) can be rewritten dimensionlessly as
\begin{equation}
    \frac{d^2\theta}{d\tau^2} = -\left(1+\frac{F_a}{T_o}\right)\theta - \frac{1}{Q}\frac{d\theta}{d\tau}
\label{EOMdimensionless}
\end{equation}
where $\tau=\Omega_o t$ is the time variable. As an example, we ask what time-dependent applied force is needed to achieve periodic motion with the particular non-sinusoidal form
\begin{equation}
    \theta(\tau)=\theta_o\left[\cos(\tau)+\epsilon\sin(2\tau)\right],
\label{eq_anharmonicmotion}
\end{equation}
where $\epsilon$ controls the amount of anharmonicity.  By direct substitution into Eq.~(\ref{EOMdimensionless}), the required applied force profile is solved to be
\begin{equation}
    F_a(\tau) = \frac{T_o}{Q}
    \frac{\tan(\tau)+6Q\epsilon\sin(\tau)-2\epsilon\frac{\cos(2\tau)}{\cos(\tau)}}{1+2\epsilon\sin(\tau)}.
\label{eq_anharmonicforce}
\end{equation}
Note that $\theta(\tau)$ becomes perfectly sinusoidal, and $F_a(\tau)$ reduces to Eq.~(\ref{Fa}), in the $\epsilon=0$ limit. For $Q=5$, the cases $\epsilon=\{0, 0.1\}$ are compared in Fig.~\ref{fig_appendix}.  For $\epsilon=0.1$ the motion appears quite sinusoidal but the force profile is markedly different from that required for perfectly sinusoidal motion.

\bibliography{BuzzerRefs}

\end{document}